\documentclass[conference]{IEEEtran} 
\IEEEoverridecommandlockouts
\usepackage{cite}
\usepackage{amsmath,amssymb,amsfonts}
\usepackage{algorithmic}
\usepackage{graphicx}
\usepackage{textcomp}
\usepackage{xcolor}
\def\BibTeX{{\rm B\kern-.05em{\sc i\kern-.025em b}\kern-.08em
    T\kern-.1667em\lower.7ex\hbox{E}\kern-.125emX}}
\begin{document}

\title{Hypersparse Network Flow Analysis of Packets with GraphBLAS
\thanks{This material is based upon work supported by the Assistant Secretary of Defense for Research and Engineering under Air Force Contract No. FA8702-15-D-0001, National Science Foundation CCF-1533644, and United States Air Force Research Laboratory and Artificial Intelligence Accelerator Cooperative Agreement Number FA8750-19-2-1000. Any opinions, findings, conclusions or recommendations expressed in this material are those of the author(s) and do not necessarily reflect the views of the Assistant Secretary of Defense for Research and Engineering, the National Science Foundation, or the United States Air Force. The U.S. Government is authorized to reproduce and distribute reprints for Government purposes notwithstanding any copyright notation herein.}
}

\author{\IEEEauthorblockN{Tyler Trigg$^1$, Chad Meiners$^1$, Sandeep Pisharody$^1$, Hayden Jananthan$^1$, Michael Jones$^1$,  Adam Michaleas$^1$,  \\ Timothy Davis$^2$, Erik Welch$^3$, William Arcand$^1$, David Bestor$^1$, William Bergeron$^1$, \\ Chansup Byun$^1$, Vijay Gadepally$^1$,  Micheal Houle$^1$, Matthew Hubbell$^1$, Anna Klein$^1$, \\ Peter Michaleas$^1$, Lauren Milechin$^1$, Julie Mullen$^1$,  Andrew Prout$^1$,  Albert Reuther$^1$,\\ Antonio Rosa$^1$, Siddharth Samsi$^1$,  Doug Stetson$^1$, Charles Yee$^1$, Jeremy Kepner$^1$
\\
\IEEEauthorblockA{$^1$MIT,  $^2$Texas A\&M, $^3$Nvidia
}}}
\maketitle

\begin{abstract}
Internet analysis is a major challenge due to the volume and rate of network traffic. In lieu of analyzing traffic as raw packets, network analysts often rely on compressed network flows (netflows) that contain the start time, stop time, source, destination, and number of packets in each direction. However, many traffic analyses benefit from temporal aggregation of multiple simultaneous netflows, which can be computationally challenging. To alleviate this concern, a novel netflow compression and resampling method has been developed leveraging GraphBLAS hyperspace traffic matrices that preserve anonymization while enabling subrange analysis. Standard multi-temporal spatial analyses are then performed on each subrange to generate detailed statistical aggregates of the source packets, source fan-out, unique links, destination fan-in, and destination packets of each subrange which can then be used for background modeling and anomaly detection. A simple file format based on GraphBLAS sparse matrices is developed for storing these statistical aggregates. This method is scale tested on the MIT SuperCloud using a 50 trillion packet netflow corpus from several hundred sites collected over several months. The resulting compression achieved is significant ($<$0.1 bit per packet) enabling extremely large netflow analyses to be stored and transported. The single node parallel performance is analyzed in terms of both processors and threads showing that a single node can perform hundreds of simultaneous analyses at over a million packets/sec (roughly equivalent to a 10 Gigabit link).
\end{abstract}


\section{Introduction}


Internet traffic analysis is crucial for billing, provisioning, forecasting, and security reasons. While analyses of raw packets was attempted \cite{paxson1999bro, roesch1999snort}, it was broadly accepted that such analyses have inherent scalability problems that restrict their deployment to lower speed links. Network Flow (netflow) \cite{claise2004cisco, estan2004building} is a compressed data format that strikes a balance between high-fidelity data and scalability. By sampling network traffic for a period of time, and aggregating measurements, netflows provide a balance between scalability and fidelity. Such netflows usually contain: Input interface port, IP source address, IP destination address, Source port number, Destination port number, Layer 3 protocol field, and Type of service along with the start and end times of the sampling window. While netflow-based traffic analysis captures the needs of billing, provisioning, and forecasting communities, it falls short of satisfying all the needs the Internet security community. 

Network analysis has emerged as important application area for the protection and improvement of the Internet.  These analyses require a significant amount of network traffic from a variety of observatories and outposts \cite{kepner2021zero, weed2022beyond}.  Historically, the data volumes, processing requirements, and privacy concerns of analyzing a significant fraction of the Internet have been prohibitive.  The North American Internet generates billions of non-video Internet packets each second \cite{Cisco2017, Cisco2018-2023}.   Novel compression, anonymization, and analysis technique are required to meet these challenges.

The GraphBLAS standard provides significant performance and compression capabilities which improve the feasibility of analyzing these volumes of data \cite{kepner2011graph, kepner2015graphs, kepner16mathematical, buluc17design, kepner2017enabling, yang2018implementing, davis18algorithm, kepner2018mathematics, davis2019algorithm, mattson2019lagraph, cailliau2019redisgraph, davis2019write, aznaveh2020parallel, brock2021introduction, pelletier2021graphblas}.   Specifically, the GraphBLAS is ideally suited for both constructing and analyzing anonymized hypersparse traffic matrices.  Prior work with the GraphBLAS has demonstrated rates of 75 billion packets per second (pps) \cite{kepner202075}, while achieving compressions of 1 bit per packet \cite{kepner2020multi}, and enabling the analysis of the largest publicly available historical archives with over 40 trillion packets \cite{kepner2021spatial}.  Analysis of anonymized hypersparse traffic matrices from a variety of sources has revealed power-law distributions \cite{kepner19hypersparse, kepner2022new}, novel scaling relations \cite{kepner2020multi, kepner2021spatial}, and inspired new models of network traffic \cite{devlin2021hybrid}.

While some raw packet corpora and datasets exist, many data taps prefer netflow, owing to the compression and scalability it offers. However, many traffic analyses for security applications benefit from temporal aggregation of multiple simultaneous netflows into hypersparse traffic matrices, which can be computationally challenging.  To solve this problem a novel netflow compression and resampling method has been developed leveraging GraphBLAS hyperspace traffic matrices that preserves anonymization.   Furthermore, standard multi-temporal spatial analyses are then performed on each subrange to generate detailed statistical aggregates of the source packets, source fan-out, unique links, destination fan-in, and destination packets of each subrange which can be used for background modeling and anomaly detection.  A simple file format based on GraphBLAS sparse matrices is developed for storing these statistical aggregates.

The outline of the rest of the paper is as follows.  Section~\ref{sec:analytics} describes the standard anonymized hierarchical hypersparse analytics for computing source packets, source fan-out, unique links, destination fan-in, and destination packets.  Next, the generation of netflow traffic matrices and the novel netflow time matrix list (TML) format is described.   Subsequently, in Section~\ref{sec:test} the netflow test data set is summarized along with the corresponding MIT SuperCloud test hardware.  Finally, Section~\ref{sec:results} captures the compression and performance results; and Section~\ref{sec:concl} captures conclusions, and directions for further work.

\section{Anonymized Hierarchical Hypersparse Analytics}
\label{sec:analytics}

\begin{figure}
\center{\includegraphics[width=1.0\columnwidth]{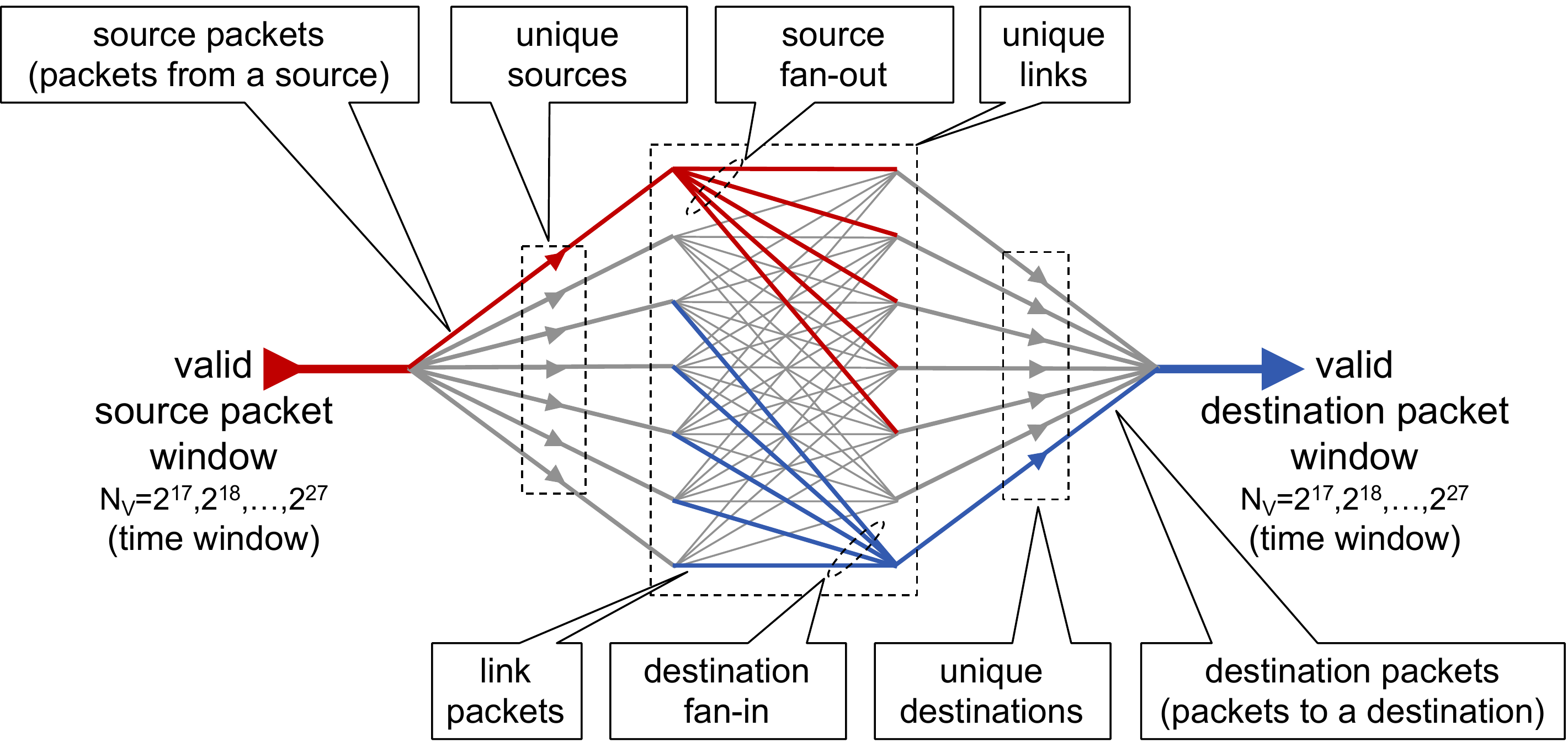}}
      	\caption{{\bf Streaming network traffic quantities.} Internet traffic streams of $N_V$ valid packets are divided into a variety of quantities for analysis: source packets, source fan-out, unique source-destination pair packets (or links), destination fan-in, and destination packets.}
      	\label{fig:NetworkDistribution}
\end{figure}

Network data must be handled with great care and privacy is a paramount concern.  A primary benefit of constructing anonymized hypersparse traffic matrices with the GraphBLAS is the efficient computation of a wide range of network quantities via matrix mathematics\cite{pisharody2021realizing}.  Figure~\ref{fig:NetworkDistribution} illustrates essential quantities found in all streaming dynamic networks. These quantities are all computable from anonymized traffic matrices created from the source and destination addresses found in Internet packet headers.

\begin{table}
\caption{Network Quantities from Traffic Matrices}
\vspace{-0.25cm}
Formulas for computing network quantities from  traffic matrix ${\bf A}_t$ at time $t$ in both summation and matrix notation. ${\bf 1}$ is a column vector of all 1's, $^{\sf T}$  is the transpose operation, and $|~|_0$ is the zero-norm that sets each nonzero value of its argument to 1\cite{karvanen2003measuring}.  These formulas are unaffected by matrix permutations and will work on anonymized data.
\begin{center}
\begin{tabular}{p{1.45in}p{0.9in}p{0.6in}}
\hline
{\bf Aggregate} & {\bf ~~~~Summation} & {\bf ~Matrix} \\
{\bf Property} & {\bf ~~~~~~Notation} & {\bf Notation} \\
\hline
Valid packets $N_V$ & $~~\sum_i ~ \sum_j ~ {\bf A}_t(i,j)$ & $~{\bf 1}^{\sf T} {\bf A}_t {\bf 1}$ \\
Unique links & $~~\sum_i ~ \sum_j |{\bf A}_t(i,j)|_0$  & ${\bf 1}^{\sf T}|{\bf A}_t|_0 {\bf 1}$ \\
Link packets from $i$ to $j$ & $~~~~~~~~~~~~~~{\bf A}_t(i,j)$ & ~~~$~{\bf A}_t$ \\
Max link packets ($d_{\rm max}$) & $~~~~~\max_{ij}{\bf A}_t(i,j)$ & $\max({\bf A}_t)$ \\
\hline
Unique sources & $~\sum_i |\sum_j ~ {\bf A}_t(i,j)|_0$  & ${\bf 1}^{\sf T}|{\bf A}_t {\bf 1}|_0$ \\
Packets from source $i$ & $~~~~~~~\sum_j ~ {\bf A}_t(i,j)$ & ~~$~~{\bf A}_t  {\bf 1}$ \\
Max source packets ($d_{\rm max}$)  & $ \max_i \sum_j ~ {\bf A}_t(i,j)$ & $\max({\bf A}_t {\bf 1})$ \\
Source fan-out from $i$ & $~~~~~~~~~~\sum_j |{\bf A}_t(i,j)|_0$  & ~~~$|{\bf A}_t|_0 {\bf 1}$ \\
Max source fan-out ($d_{\rm max}$) & $ \max_i \sum_j |{\bf A}_t(i,j)|_0$  & $\max(|{\bf A}_t|_0 {\bf 1})$ \\
\hline
Unique destinations & $~\sum_j |\sum_i ~ {\bf A}_t(i,j)|_0$ & $|{\bf 1}^{\sf T} {\bf A}_t|_0 {\bf 1}$ \\
Destination packets to $j$ & $~~~~~~~\sum_i ~ {\bf A}_t(i,j)$ & ${\bf 1}^{\sf T}|{\bf A}_t|_0$ \\
Max destination packets ($d_{\rm max}$) & $ \max_j \sum_i ~ {\bf A}_t(i,j)$ & $\max({\bf 1}^{\sf T}|{\bf A}_t|_0)$ \\
Destination fan-in to $j$ & $~~~~~~~~~~\sum_i |{\bf A}_t(i,j)|_0$ & ${\bf 1}^{\sf T}~{\bf A}_t$ \\
Max destination fan-in ($d_{\rm max}$) & $ \max_j \sum_i |{\bf A}_t(i,j)|_0$ & $\max({\bf 1}^{\sf T}~{\bf A}_t)$ \\
\hline
\end{tabular}
\end{center}
\label{tab:Aggregates}
\end{table}%

The network quantities depicted in Figure~\ref{fig:NetworkDistribution} are computable from anonymized origin-destination traffic  matrices that are widely used to represent network traffic \cite{soule2004identify, zhang2005estimating, mucha2010community, tune2013internet}.   To reduce statistical fluctuations, the streaming data should be partitioned such that any chosen time window all data sets have the same number of valid packets \cite{kepner19streaming}.  At a given time $t$, $N_V$ consecutive valid packets are aggregated from the traffic into a hypersparse matrix ${\bf A}_t$, where ${\bf A}_t(i,j)$ is the number of valid packets between the source $i$ and destination $j$. The sum of all the entries in ${\bf A}_t$ is equal to $N_V$
$$
    \sum_{i,j} {\bf A}_t(i,j) = N_V
$$
Constant packet, variable time samples simplify the statistical analysis of the heavy-tail distributions commonly found in network traffic quantities \cite{kepner19hypersparse, nair2020fundamentals, kepner2022new}.  All the network quantities depicted in Figure~\ref{fig:NetworkDistribution} can be readily computed from ${\bf A}_t$ using the formulas listed in Table~\ref{tab:Aggregates}.  Because matrix operations are generally invariant to permutation (reordering of the rows and columns), these quantities can readily be computed from anonymized data.  

The contiguous nature of these data allows the exploration of a wide range of packet windows from $N_V = 2^{17}$ (sub-second) to $N_V = 2^{27}$ (minutes), providing a unique view into how network quantities depend upon time.  These observations provide new insights into normal network background traffic that could be used for anomaly detection, AI feature engineering, polystore index learning, and testing theoretical models of streaming networks \cite{elmore2015demonstration, kraska18case, do20classifying}.

\begin{figure}
\center{\includegraphics[width=1.0\columnwidth]{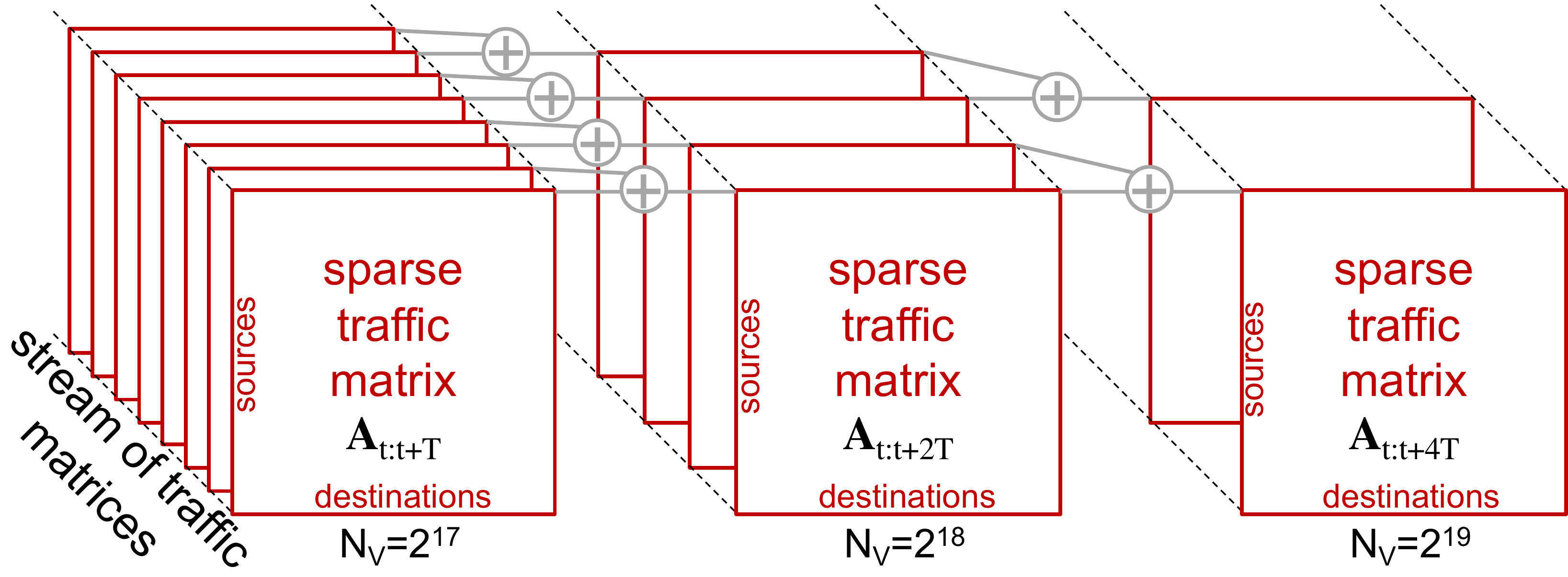}}
      	\caption{{\bf Multi-temporal streaming traffic matrices.} Efficient computation of network quantities on multiple time scales can be achieved by hierarchically aggregating data in different time windows.}
      	\label{fig:MultiTemporalMatrix}
\end{figure}

Network traffic is dynamic and exhibits varying behavior on a wide range of time scales.  A given packet window size $N_V$ will be sensitive to phenomena on its corresponding timescale.  Determining how network quantities scale with $N_V$ provides insight into the temporal behavior of network traffic.   Efficient computation of network quantities on multiple time scales can be achieved by hierarchically aggregating data in different time windows \cite{kepner19streaming}.  Figure~\ref{fig:MultiTemporalMatrix} illustrates a binary aggregation of  different streaming traffic matrices.   Computing each quantity at each hierarchy level eliminates redundant computations that would be performed if each packet window was computed separately.  Hierarchy also ensures that most computations are performed on smaller matrices residing in faster memory.  Correlations among the matrices mean  that adding two matrices each with $N_V$ entries results in a matrix with fewer than $2N_V$ entries, reducing the relative number of operations as the matrices grow.

One of the important capabilities of the SuiteSparse GraphBLAS library is direct support of hypersparse matrices where the number of nonzero entries is significantly less than either dimensions of the matrix.  If the packet source and destination identifiers are drawn from a large numeric range, such as those used in the Internet protocol, then a hypersparse representation of ${\bf A}_t$ eliminates the need to keep track of additional indices and can significantly accelerate the computations \cite{kepner202075}.

\begin{figure}
\center{\includegraphics[width=0.65\columnwidth]{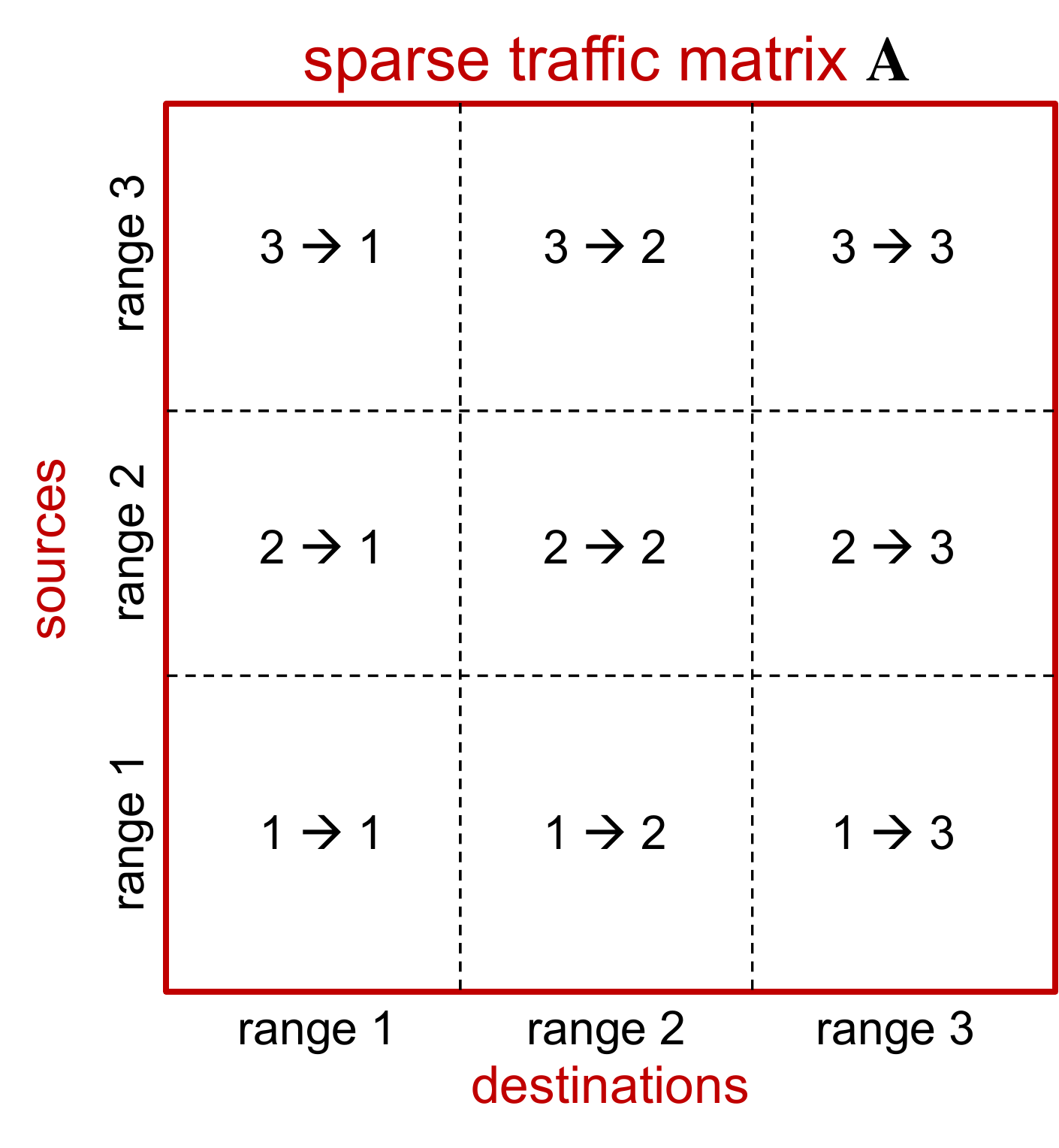}}
      	\caption{{\bf Traffic matrix ranges.}   Traffic matrices can be divided into subranges of specific sources and destinations for analysis, which can be essential if different subranges have different statistical distributions.}
      	\label{fig:SubRangeTrafficMatrix}
\end{figure}

It is common to filter the packets down to a valid set for  any particular analysis.  Such filters may limit particular source ranges, destinations ranges, protocols, and time windows.  Anonymized data can be analyzed by subranges of IPs using simple matrix multiplication.  For a given subrange represented by an anonymized hypersparse diagonal matrix ${\bf A}_r$, where ${\bf A}_r(i,i) = 1$ implies  source/destination $i$ is in the range, the traffic within the subrange can be computed via: ${\bf A}_r {\bf A}_t  {\bf A}_r$. Likewise, for additional privacy guarantees that can be implemented at the  edge, the same method can be used to exclude a range of data from the traffic matrix
$$
     {\bf A}_t - {\bf A}_r {\bf A}_t  {\bf A}_r
$$ 
In this work, three source and destination subranges are used corresponding to non-routable traffic, bogon traffic, and all other traffic (see Figure~\ref{fig:SubRangeTrafficMatrix}). All network quantities are computed on the entire traffic matrix and each of the 9 smaller traffic matrices. 

\section{Netflow Traffic Matrices}
\label{sec:matrix}

Use of compressed netflow data to generate matrices of exactly $N_V$ packets needs time normalization to align with the constant packet, variable time matrix sizing approach described above. With time normalization, the aggregated netflow records are converted into a series of timestamped flows that enable the grouping of netflows to build network matrices with of fixed size. The time intervals associated with flow entries are normalized into Unix epoch time with a resolution of seconds. If a netflow record lasts for longer than one second, an even data transfer rate is assumed, and the packet counts ($pkts$) are divided evenly to each second within the interval.

For flow records whose number of packets ($pkts$) do not divide evenly by the number of 1-second time bins ($tbins$), the number of bins that will contain an extra packet $extra$ is given by 
 \begin{equation*}
 extra = pkts - \left(\lfloor\frac{pkts}{tbins}\rfloor\times tbins\right)
 \end{equation*}

If $extra \leq 2$, the extra bins are allocated to the first and last bin in the interval.  Otherwise, the remaining extra bins are shuffled into the bins between the first and last bin. This creates an even distribution of flows across the time interval.  All packets at a given time bin are then aggregate into traffic matrix for that time bin.

\begin{figure*}
\center{\includegraphics[width=1.3\columnwidth]{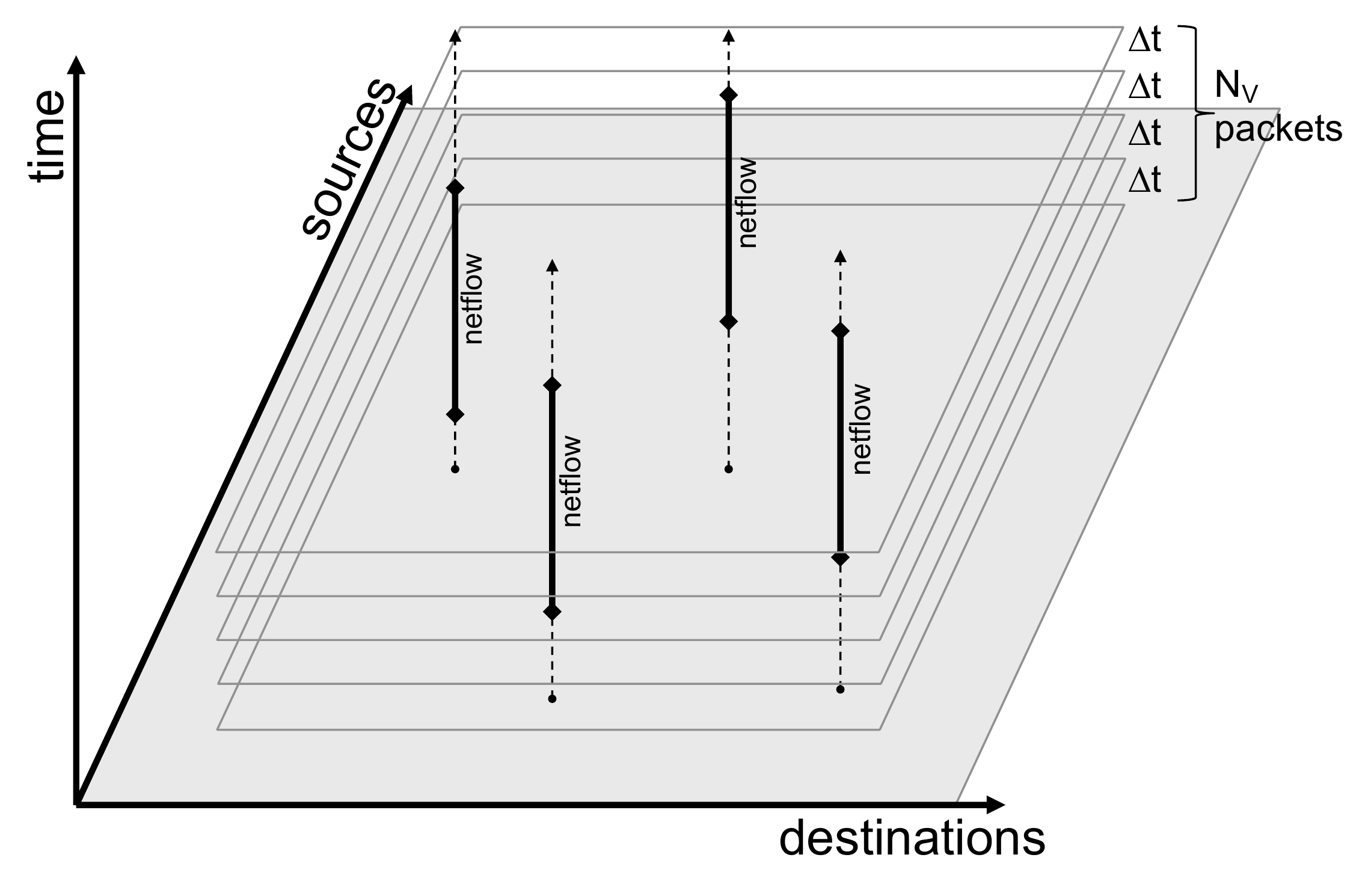}}
      	\caption{{\bf Netflow traffic matrices.} The Internet can be viewed as a hypersparse traffic matrix where each point represents communication between a source and a destination.  Network flows (netflows) are vertical lines corresponding to a start time, stop time, source, destination, and number of packets in each direction.  Time sliced traffic matrices can be constructed by spreading the packets uniformly across time windows of width $\Delta t$.  These traffic matrices can then be summed to into aggregated traffic matrices with exactly $N_V$ valid packets for subsequent analyses.}
      	\label{fig:FlowSampling}
\end{figure*}

The representation of network flow in a matrix coordinate space opens up the possibility of further compressing netflow. The Time Compressed Matrices List (TML) format leverages the matrix coordinate space, and only stores the delta of coordinates between time-adjacent matrices, as show in Figure~\ref{fig:FlowSampling}. Using a canonical ordering of the current coordinate space (source-destination IP pairings), the deltas can then be used to reconstruct the original matrix in a lossless fashion.

TML representation of a netflow is a stream of 5-vectors $\langle(t_0,s_0,D_0,I_0,V_0), ..., (t_n,s_n,D_n,I_n,V_n)\rangle$ where $t_0$ to $t_n$ are signed 64-bits Unix epoch timestamps; $s_0$ to $s_n$ are an encoded pair of 32-bit signed delete-insert count; $D_0$ to $D_n$ are the list of deleted coordinates, $I_0$ to $I_n$ are the list of inserted coordinates, and $V_0$ to $V_n$ are the list of canonically ordered coordinates associated with current coordinate set.

\section{Test Data and Hardware}
\label{sec:test}

The test data set comprises of nearly $100$~TB of netflow data with over 50 trillion packets from many different locations for a period of several months. The traffic patterns are representative of a large enterprise environments with bidirectional traffic at the collection points. The collection points also covered a range of traffic rates.

The  analysis  code was implemented using Python GraphBLAS bindings with the pPython parallel library \cite{Byun2022}.  A typical run could be launched in a few seconds using the MIT SuperCloud triples-mode hierarchical launching system \cite{8547629}.  The launch parameters were [Nnodes Nprocess Nthread], corresponding to Nnodes nodes, Nprocess Python processes per node, and Nthread OpenMP threads per process.  On each node, each of the Nprocess processes and their corresponding Nthread threads were pinned to adjacent cores to minimize interprocess contention and maximize cache locality for the GraphBLAS OpenMP threads \cite{byun2019optimizing}.  Within each Python process, the underlying GraphBLAS OpenMP parallelism is used.  At the end of the processing the results were aggregated using asynchronous file-based messaging \cite{byun2019large}. Triples mode makes it easy to explore horizontal scaling  across nodes, vertical scaling by examining combinations of processes and threads on a node, and temporal scaling by running on diverse hardware from different eras.

The computing hardware consists of four different types of nodes acquired over a decade (see Table~\ref{tab:HardwareTable}).  The nodes are all multicore x86 compatible with comparable total memory.  The MIT SuperCloud maintains the same modern software across all nodes, which allows for direct comparison of hardware performance differences.

\begin{table}
\caption{Computer Hardware Specifications}
\vspace{-0.25cm}
MIT SuperCloud maintains a diverse set of hardware running an identical modern software stack providing an unique platform for comparing performance over different eras.
\begin{center}
\includegraphics[width=\columnwidth]{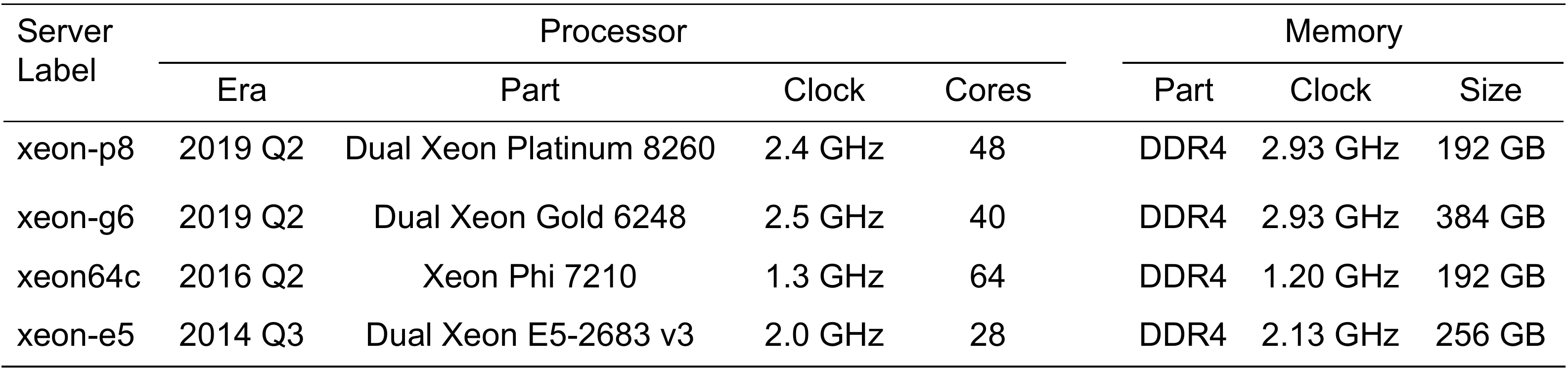}
\end{center}
\label{tab:HardwareTable}
\end{table}

\section{Results}
\label{sec:results}

\begin{figure}
\center{\includegraphics[width=1.0\columnwidth]{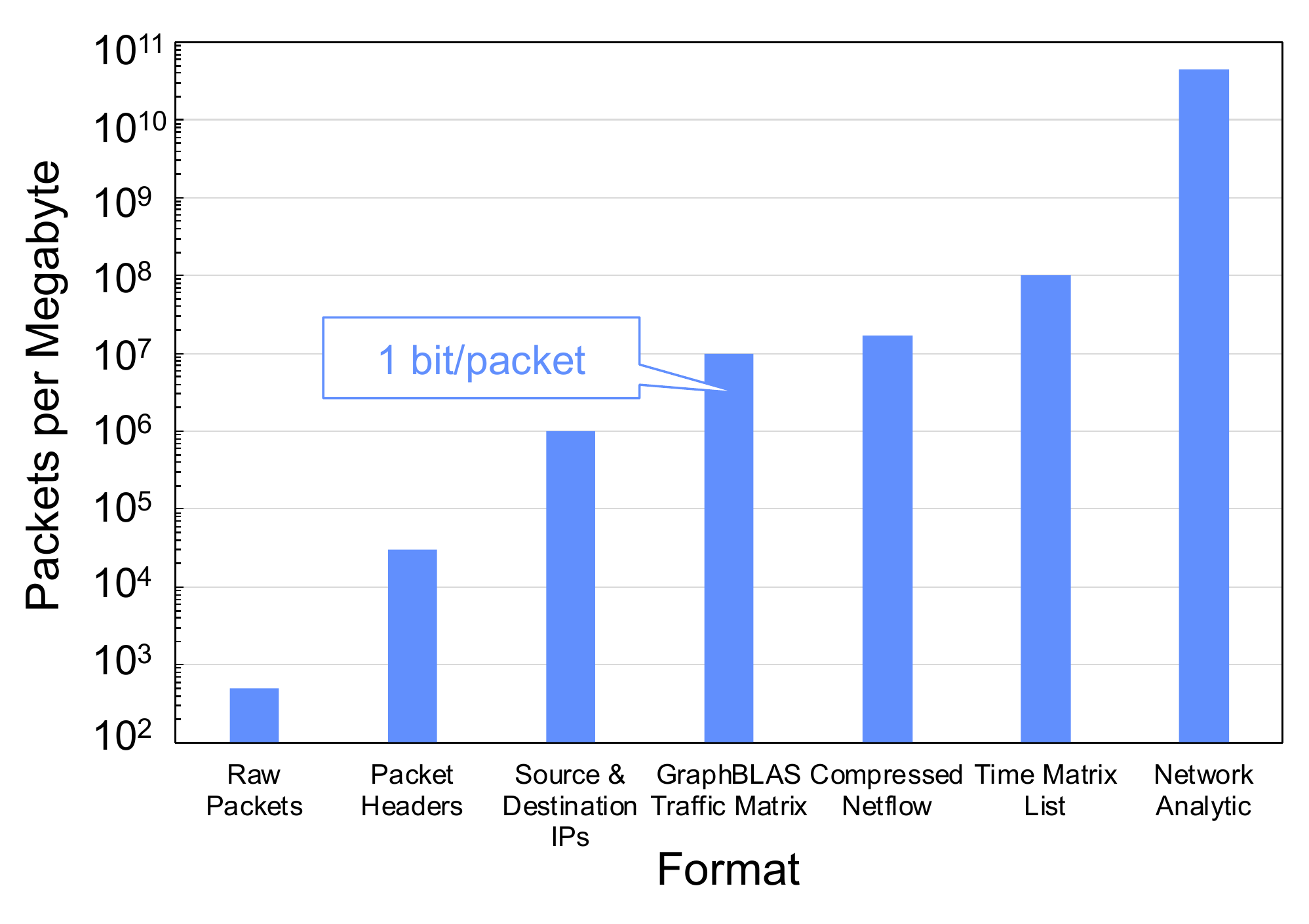}}
      	\caption{{\bf Estimated file format sizes.}  Number of packets per megabyte for different file formats in the analysis pipeline.  The time matrix list (TML) and resulting network analyses provide $10^5$x to $10^7$x compression over the raw packets.}
      	\label{fig:compression}
\end{figure}

The use of GraphBLAS to represent network traffic matrices has provided tremendous compression results \cite{kepner2021zero}. Netflow representation of raw data provides variable compression results, depending on the time intervals used for aggregation. Aggregating one hour of raw traffic into a single netflow will provide much greater compression than aggregating one minute of raw traffic - with the gain in compression being offset by the loss in data fidelity. The use of TML format as described in Section~\ref{sec:matrix} to represent netflow data provides additional compression without any further loss in data fidelity. Figure~\ref{fig:compression} plots the number of packets per megabyte for different file formats in the analysis pipeline, with the TML format and resulting network analyses providing $10^5$x to $10^7$x compression over the raw packets. For the dataset that was analyzed, the TML representation of netflow achieved significant ($93\%$) compression on the $100$ TB dataset that was used for analysis.

\begin{figure*}[]
\centering
\includegraphics[width=\columnwidth]{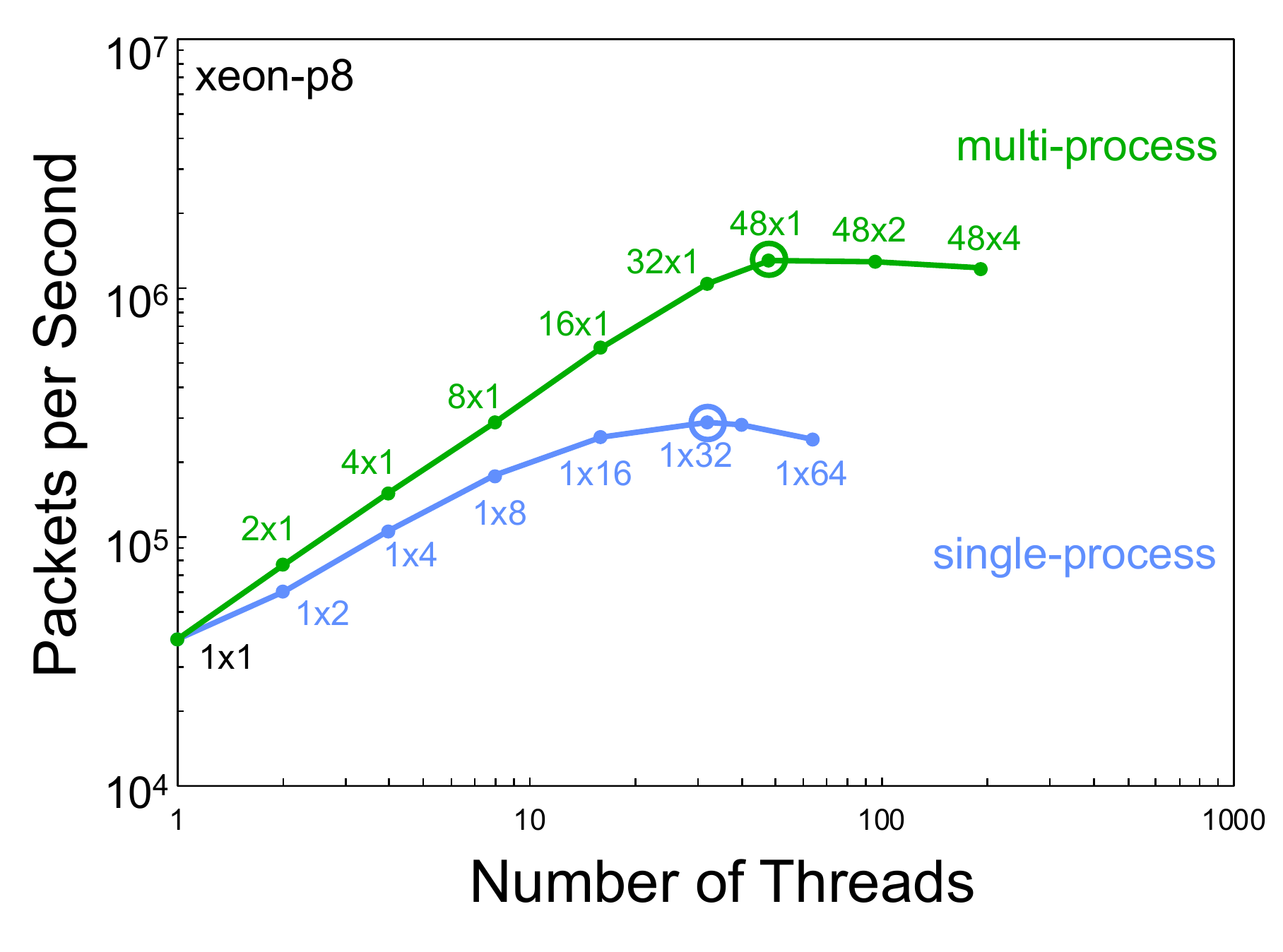}
\includegraphics[width=\columnwidth]{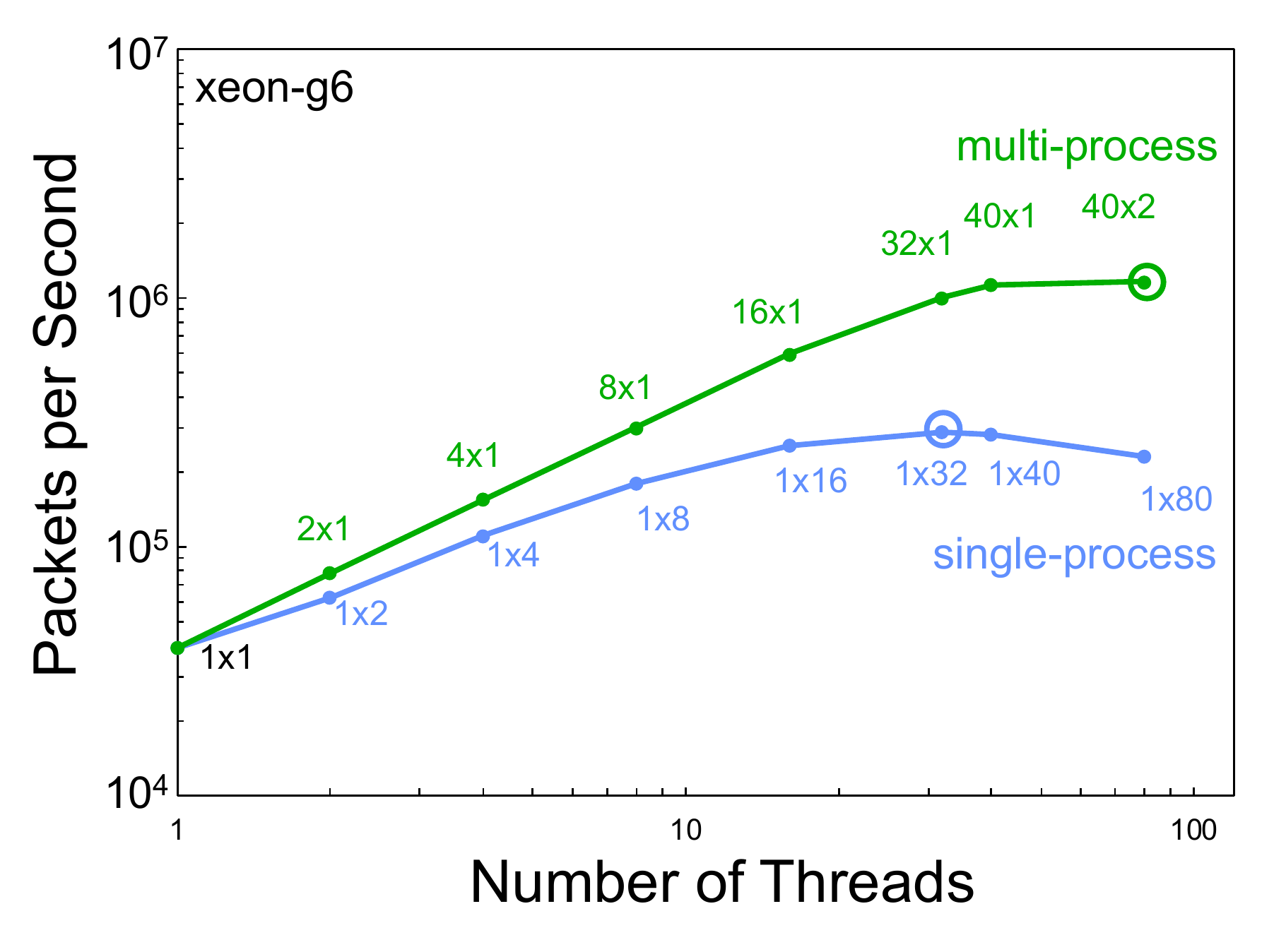}
\includegraphics[width=\columnwidth]{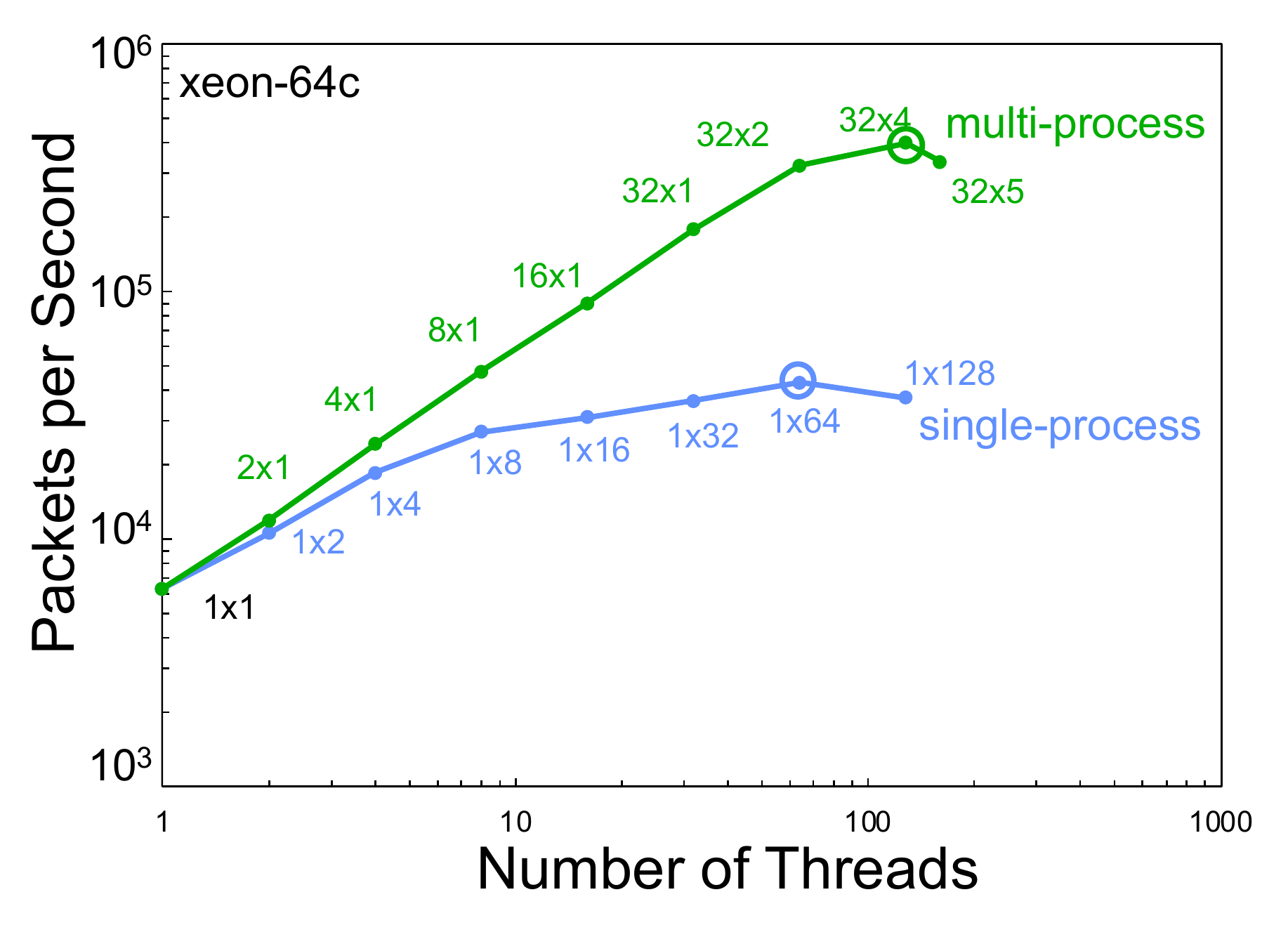}
\includegraphics[width=\columnwidth]{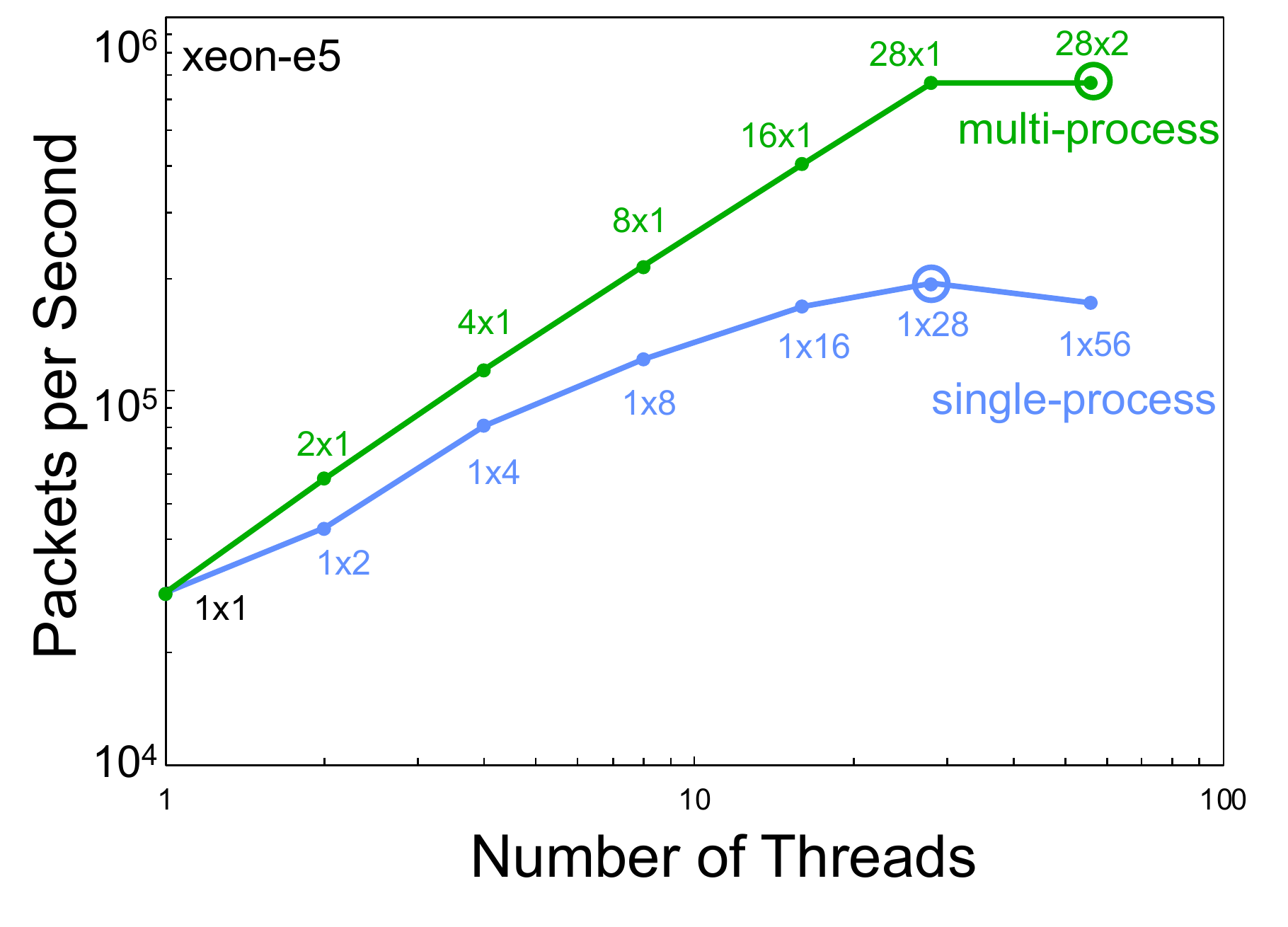}
\caption{Single node performance with different configurations of processes and threads on a xeon-p8 (upper left), xeon-g6 (upper right), xeon64c (bottom left), and xeon-e5 (bottom right).  Each data point is labeled: (\#processes)x(\#threads/process). The maximum performance for each case is denoted by a circle.}
\label{fig:SingleNodeRate}
\end{figure*}

The benchmarking explores how different numbers of GraphBLAS processes and threads perform on different multicore compute nodes.  This type of benchmarking is generally useful in most projects as it allows the determination of the best combination processing and threads prior to significant computation.  The number of processes and threads used for a given benchmark is denoted by $N_{process}$ $\times$ $N_{threads}$ whose product is equal to total number threads used in the computation.  The GraphBLAS computation described in the previous section was repeated for two sets of parameters: single-process and multi-process.  In the single-process case the parameters tested are
$$
  1{\times}1, 1{\times}2, 1{\times}4, 1{\times}8, ...  
$$
In the multi-process case the parameters tested are
$$
  1{\times}1, 2{\times}1, 4{\times}1, 8{\times}1, ...  
$$
Figure~\ref{fig:SingleNodeRate} shows the single node performance using different numbers of processes and threads for the different servers listed in Table~\ref{tab:HardwareTable}.  In all cases, the multi-process scaling provided greater aggregate performance and the single-process scaling provided a maximum of a 4x speedup over $1{\times}1$ case.

\section{Conclusions and Future Work}
\label{sec:concl}

Analysis of traffic on the Internet has long been a major challenge due to the volume and rate of network traffic. Recent advances, combining the use of hypersparse matrix representation of network traffic coupled with GraphBLAS is proving to be a key enabler of privacy-preserving network analytics. The effort documented in this paper further advances the field by demonstrating the ability to use compressed network flows (netflows) in lieu of raw packet streams to conduct similar analytics. Further, it introduces a novel netflow compression and resampling method that preserves anonymization while enabling subrange analysis. This method is scale tested on the MIT SuperCloud using a 50 trillion packet netflow corpus from several hundred sites collected over several months.  The resulting compression achieved is significant ($<$0.1 bit per packet) enabling extremely large netflow analyses to be stored and transported.  The single node parallel performance is analyzed in terms of both processors and threads showing that a single node can perform hundreds of simultaneous analyses at over a million packets/sec (roughly equivalent to a 10 Gigabit link).

The success in the use of netflow in place of raw packets to conduct analyses opens up a number of potential pathways for future work; amongst which are cross-correlation of netflow data from different observatories and outposts, comparing gain/loss for using different netflow aggregation windows, and exploring the gain/loss using expected utility theory and prospect theory. Further analysis needs to be done to calibrate AI algorithms for classification of background traffic to use netflow.

\section*{Acknowledgments}

The authors wish to acknowledge the following individuals for their contributions and support: Bob Bond, Stephen Buckley, Ronisha Carter, Cary Conrad, Alan Edelman, Tucker Hamilton, Jeff Gottschalk, Chris Hill, Mike Kanaan, Tim Kraska, Charles Leiserson, Mimi McClure, Kyle McAlpin, Joseph McDonald, Sandy Pentland, Heidi Perry, Christian Prothmann, John Radovan, Steve Rejto, Daniela Rus, Matthew Weiss, Marc Zissman.

\bibliographystyle{ieeetr}
\bibliography{HyperNetFlow}

\end{document}